\documentclass[12pt, notitlepage]{article}
\pdfoutput=1
\usepackage{scicite}
\usepackage{amsmath}
\usepackage{amsfonts}
\usepackage{amssymb}
\usepackage[dvipdfmx]{graphicx}
\usepackage{txfonts}
\usepackage[margin=1in, letterpaper]{geometry}
\usepackage{bm}
\usepackage{color}

\newenvironment{sciabstract}{%
\begin{quote} \bf}
{\end{quote}}

\begin{document}
\baselineskip 24pt
\title{\textbf{Interface-driven topological Hall effect in SrRuO\bm{$_3$}-SrIrO\bm{$_3$} bilayer}}
\author
{J.~Matsuno,$^{1\ast}$ N.~Ogawa,$^{1}$ K. Yasuda,$^{2}$ F.~Kagawa,$^{1}$ W.~Koshibae,$^{1}$\\
N.~Nagaosa,$^{1,2}$ Y.~Tokura,$^{1,2}$ M.~Kawasaki,$^{1,2}$ \\
\\
\normalsize{$^{1}$RIKEN Center for Emergent Matter Science (CEMS), Saitama 351-0198, Japan}\\
\normalsize{$^{2}$Department of Applied Physics, University of Tokyo, Tokyo 113-8656, Japan}\\
\normalsize{$^\ast$Corresponding author. E-mail: matsuno@riken.jp.}
}
\maketitle

\begin{sciabstract}
Electron transport coupled with magnetism has attracted attention over the years as exemplified in 
anomalous Hall effect due to a Berry phase in momentum space.
Another type of unconventional Hall effect -- topological Hall effect, originating from the real-space Berry phase,
has recently become of great importance in the context of magnetic skyrmions.
We have observed topological Hall effect in bilayers consisting of ferromagnetic SrRuO$_3$ and paramagnetic SrIrO$_3$ over a wide region of both temperature and magnetic field.
The topological term rapidly decreases with the thickness of SrRuO$_3$, ending up with the complete disappearance at 7 unit cells of SrRuO$_3$.
Combined with model calculation, we concluded that the topological Hall effect is driven by interface Dzyaloshinskii-Moriya interaction, which is caused by both the broken inversion symmetry and the strong spin-orbit coupling of SrIrO$_3$.
Such interaction is expected to realize the N\'{e}el-type magnetic skyrmion, of which size is estimated to be $\sim$10~nm from the magnitude of topological Hall resistivity.
The results established that the high-quality oxide interface enables us to tune the chirality of the system; this can be a step towards the future topological electronics.
\end{sciabstract}

\newpage
\section*{Introduction}

In the past several decades, electron transport intertwined with magnetism has been a focus of intensive research for its basic scientific importance as well as its possible technological applications.
Anomalous Hall effect (AHE), which is driven by magnetization ($M$) in ferromagnets, is one of the phenomena of interest~\cite{Nagaosa2010, Ye1999, Onoda2004, Bruno2004}.
While ordinary Hall effect is a consequence of Lorentz force and hence is proportional to magnetic field ($H$), the origin of intrinsic AHE has been clarified to be a Berry phase in momentum space in most cases. 
The other type of unconventional Hall effect which is proportional to neither $H$ nor $M$ has recently been found in a pyrochlore ferromagnet Nd$_2$Mo$_2$O$_7$;
the Hall effect has been originating from scalar spin chirality [$\chi_{jik} = \mbox{\boldmath{$S$}}_i \cdot (\mbox{\boldmath{$S$}}_j \times \mbox{\boldmath{$S$}}_k)$], which is generated by the non-coplanar configuration of Mo spins~\cite{Taguchi2001}.
This can be attributed to the Berry phase in real space and therefore is termed topological Hall effect (THE). 
Now it is widely known that THE has been observed as well in metallic magnets that host magnetic skyrmions (see Fig.~1B), which are topologically protected nanometer-sized spin swirling textures endowed with scalar spin chirality~\cite{Rossler2006,Muhlbauer2009,Yu2010,Yu2011,Nagaosa2013}.
Among them, particularly important is the one formed by Dzyaloshinskii-Moriya (DM) interaction, giving rise to smaller skyrmion size of 5--100~nm; skyrmion-driven THE has been reported for metallic magnets with chiral crystal structure such as B20 compounds~\cite{Neubauer2009,Kanazawa2011,Yokouchi2014}, demonstrating that THE is a promising tool for probing the skyrmion.

Considering that DM interaction arises from spin-orbit coupling combined with broken inversion symmetry, it is possible to artificially introduce DM interaction at surface/interface.
This concept is indeed verified at the interface between 3$d$ ferromagnetic metal (Mn, Fe, and Co) and 5$d$ paramagnetic metal (W and Ir) by surface-sensitive techniques such as spin-polarized scanning tunneling microscopy and spin-polarized low energy electron microscopy; chirality of surface magnetism has been reported~\cite{Bode2007,Romming2015,Chen2013}. 
In this paper, we investigate the interface DM interaction by measurements of THE. 
Given that high-quality epitaxial interfaces are extensively investigated in various perovskite-type transition-metal oxides during the last decade~\cite{Hwang2012}, we combined two transition-metal oxides in the form of epitaxial bilayers of perovskite:
one is SrRuO$_3$, a well-known itinerant ferromagnet with a Curie temperature of $\sim$160~K~\cite{Koster2012}. 
The other is SrIrO$_3$, a paramagnetic semimetal that has been lately clarified to host 5$d$ electrons with strong-spin orbit coupling~\cite{Matsuno2015}. 
The interface between SrRuO$_3$ and SrIrO$_3$ offers an ideal arena to search for the DM interaction due to the broken inversion symmetry for the following reasons:
(i) There is no charge transfer due to the polar catastrophe~\cite{Nakagawa2006} since it contains common $A$ site ion (Sr$^{2+}$) with stable-valence $B$ site ions, Ru$^{4+}$ and Ir$^{4+}$. 
(ii) The lattice mismatch at the interface is quite small (0.48\%); the averaged lattice constants are 0.3923~\cite{Jones1989} and 0.3942~nm~\cite{Zhao2008}, respectively.
The studied structure is schematically depicted in Fig.~1C.
We used the bilayer consisting of $m$ unit cells of SrRuO$_3$ and 2 unit cells of SrIrO$_3$.
At the interface we can expect the finite DM vector pointing the in-plane direction, which may give rise to N\'{e}el-type magnetic skyrmion~\cite{Fert2013,Heinze2011,Jiang2015,Moreau-Luchaire2016,Wakatsuki2015}.
Indeed, we have observed that topological Hall effect only when $m$ is as small as 4--6, suggesting that it is derived from interface DM intearction.

\section*{Results}

\subsection*{Basic physical properties of the SrRuO$_3$-SrIrO$_3$ bilayers}

Epitaxial bilayers consisting of SrRuO$_3$ and SrIrO$_3$ were deposited on SrTiO$_3$(001) substrates by pulsed laser deposition.
Interface quality was examined by an atomically resolved HAADF (high-angle-angular-dark-
field)-STEM (scanning transmission electron microscopy) with enhanced atomic number contrast, as shown in
Fig.~1C; two layers of Ir atoms indicated by the brightest spots are accurately aligned in the [001] plane, manifesting the abrupt interface. 
The bilayer samples were further characterized by transport and magnetic measurements as displayed in Fig.~1A.
The resistivity decreases systematically with $m$.
While the samples with $m \geq 5$ have resistivity less than 1~m$\Omega$cm and metallic temperature dependence, the $m=4$ bilayer has particularly higher resistivity with a small upturn below 50~K.
These indicate that the electrons in the bilayers tend to be localized with decreasing $m$, reaching to the nearly insulating state at $m=4$.
The magnetoresistance (MR) also suggests that the ferromagnetic state is relatively destabilized with decreasing $m$ as can be seen in the systematic shift of the MR peaks that are known to correspond to the Curie temperature ($T_\mathrm{C}$).
The magnetization clearly evidences that both the saturated moment and the $T_\mathrm{C}$ are suppressed for smaller $m$; the $T_\mathrm{C}$ almost reaches down to 90~K at $m=4$.
This is naturally expected because the ferromagnetism of SrRuO$_3$ is driven by its itinerant properties.
The same trend has been reported for ultrathin SrRuO$_3$ films~\cite{Xia2009}.
In order to see more closely the $m$-dependent evolution of the transport and magnetic properties, we measured AHE (details of AHE as a function of magnetic field are discussed later).
Anomalous Hall conductance is plotted as a function of magnetization in Fig.~1D, demonstrating both the sign change and the good scaling with variation of $M$.
The preceding studies reported on the same scaling behavior, which have been attributed to the actions of momentum-space monopoles in the band structure of SrRuO$_3$~\cite{Fang2003,Mathieu2004}.
Therefore the observed scaling relationship indicates that the magnetic transport properties of the bilayers are governed by those of SrRuO$_3$.
The systematics found in resistivity, magnetoresistance and AHE reveals that the samples are precisely controlled by $m$, the thickness of SrRuO$_3$.

\subsection*{Anomaly in Hall resistivity: topological Hall effect}

We observed the clear anomaly in the Hall resistivity only in case of small $m$.
Figure~2A shows the Hall resistivity of the bilayers as a function of magnetic field.  
In $m=4$, we can clearly see the unconventional behavior of the Hall resistivity below 60~K while the overall lineshape is dominated by the positive AHE.
At 5~K, for example, the red curve indicates the hump structure between 0.8 and 2.1~T with increasing the magnetic field.
When we reverse the sweep direction, in contrast, the data drawn in blue is monotonic in the same field range.
In general, Hall resistivity is expressed by
\[ \rho_\mathrm{H} = R_0 H + R_S M + \rho_H^T, \]
where the first, second, and third terms denote the ordinary, anomalous, and topological Hall effects, respectively.
Here we neglect the first term that is already subtracted from the data in Fig.~2. 
The observed non-monotonic hump structure can never be attributed to the magnetization ($M$-$H$) curve; we can assign the structure to the third term, topological Hall effect.
Increasing the thickness $m$ to 5, we can still discern the similar peak structure.
To more clearly demonstrate it, the magnified view of the detailed temperature dependence is presented in Fig.~2B.
At lower temperatures of 60~K, $\rho_{H}$ drawn in red consists of negative AHE and an additional peak at around 0.15~T.
AHE goes across zero at 70 K, at which Hall resistivity accidentally provides the genuine topological Hall component; it represents its maximum at 0.1~T.
At 80 K and 90 K, we can detect THE with positive AHE, giving rise to the similar lineshape with $m=4$.
In $m=6$, a very tiny THE is discerned whereas it is indistinguishable in the scale of Fig.~2A.
Eventually we do not observe any unconventional feature for $m=7$.

In order to precisely evaluate THE, we separated AHE and THE by measuring the Kerr rotation
angle; the Kerr signal magnitude is anticipated to be proportional to $M$ and hence AHE as a function of magnetic field at a fixed temperature (see Section A in the Supplementary Materials for comparison between the Kerr rotation angle and the magnetization).
The representative data set of $m=5$ at 80~K is plotted in Fig.~2C.
At high magnetic field region where the magnetization is saturated, all the spins align ferromagnetically, leading to the absence of scalar spin chirality; the Hall resistivity is attributed only to AHE.
Then we can fit the Hall resistivity by using the Kerr rotation angle to obtain AHE.
Figure~2C establishes that the fitting is quite well performed, illustrating that we can obtain THE by subtracting AHE from the total Hall resistivity.
The resultant THE in Fig.~2C has a very similar lineshape with that of the the 70~K-data in Fig.~2B discussed above, indicating that the subtraction procedure works well.
Increasing the field from $-9$~T to 0~T, the magnetic state is totally dominated by ferromagnetic state with negative magnetization.
This corresponds to the observed finite AHE and the negligible THE.
With further increase of the magnetic field from 0~T to 0.8~T, THE abruptly takes a peak at 0.06~T and gradually decreases to zero at 0.4~T, which coincides with the field at which the hysteresis in $M$-$H$ curve closes.
This suggests that some specific spin structure with finite scaler spin chirality is induced when the ferromagnetic spins begin to be reversed.
The simultaneous observation of the hysteresis and the THE indicates a coexistence between the ferromagnetic phase and the phase with scalar spin chirality.
We also note that the scalar spin chirality was observed only by the transport property through emergent magnetic field, i.e.,\ a fictitious magnetic field derived from the real-space Berry phase, while the chirality only marginally affects the magnetization.

We applied the same procedure to all the data shown in Fig.~2A to obtain the topological Hall term as functions of both $T$ and $H$.
As clearly exemplified in Fig.~2D for $m=5$, a sign of THE is always positive, irrespective of the sign change of AHE at 70~K~\cite{sign}.
This indicates that the unconventional THE has a totally different origin from AHE.
Instead, we confirm again that THE is driven by magnetization reversal process since the peak position of 
$\rho_H^T$ ($H_p$) scales quite well with coercive field ($H_c$).
We also note that the topological Hall term is observed in the wide range of $T$-$H$ plane. 
The most plausible spin-chiral structure responsible for the THE is the magnetic skyrmion.
In the bulk B20 compounds, the lattice form of skyrmions (Bloch-type, Fig~1B) was found in a very narrow $T$-$H$ window close to the Curie temperature~\cite{Muhlbauer2009}.
In contrast, thin films of B20 compounds are reported to stabilize the skyrmion, which is discussed in terms of the film thickness relative to the skyrmion size~\cite{Yu2011,Yokouchi2014}.
In our case of ultrathin SrRuO$_3$ films with roughly 2~nm thickness, we can reasonably expect the stability of the two dimensional skyrmion, which is consistent with the observed wide-range THE.

\subsection*{Ferromagnet-thickness dependence of topological Hall effect}

We now clarify the importance of interface from the $m$-dependence of the topological Hall term.
Figure~3A plots $\rho_H^T(m, T)$, the maximum value of the topological Hall term at $H=H_p$, signifying that it decreases with $m$ ending up the complete disappearance at $m=7$. 
This $m$ dependence can be qualitatively explained by assuming the DM vector only at the interface.
In order to realize spin texture with finite spin chirality, all the spins of SrRuO$_3$ through the thickness should be twisted by the interface DM interaction; the energy cost for twisting the same angle linearly scales with $m$, i.e., volume of the ferromagnet as shown in schematics in bottom panel of Fig.~3A.
In other words, effective DM interaction ($\equiv D_\mathrm{eff}$) is expected to decrease with $m$.
For more precise understanding, we performed the following two-step analysis: i) to investigate $D_\mathrm{eff}$ as a function of $m$ and (ii) to estimate the energetics of two-dimensional skyrmion with $D_\mathrm{eff}$.

In order to elucidate the $m$ dependene of the effective DM interaction, we numerically examined  a single skyrmion stability in a multilayer system modelled by the following Hamiltonian (see Section B in the Supplementary Materials for details of the calculation.):
\[
H = -J \sum_{<l\bm{i}l'\bm{i}'>} \bm{n}_{l,\bm{i}} \cdot \bm{n}_{l,\bm{i}'} 
    +D \sum_{\bm{i}}\left[\hat{\bm{y}}\cdot\bigl(\bm{n}_{1,\bm{i}}\times\bm{n}_{1,\bm{i}+\hat{\bm{x}}}\bigr)
     - \hat{\bm{x}} \cdot \bigl(\bm{n}_{1,\bm{i}} \times \bm{n}_{1,\bm{i}+\hat{\bm{y}}}\bigr) \right]
     - h\sum_{l=1}^m\sum_{\bm{i}} n^z_{l,\bm{i}}\,,
\label{Ham}
\]
where $J$ is the ferromagnetic coupling constant and $D$ is the DM interaction only on the first layer ($l=1$).
The normalized magnetic moment at the site $\bm{i}$ on the layer $l$ is denoted as $\bm{n}_{l,\bm{i}}$. 
The unit vectors $\hat{\bm{x}}$ and $\hat{\bm{y}}$ define the two-dimensional square lattice on a layer.
The last term represents the Zeeman energy with external magnetic field $h$ perpendicular to the multilayer.  
Control parameters are $D$ and total number of layers $m$, while we fix $J=1.0$ and  $h=0.01$.
Through the whole parameter range, we obtained  three types of magnetic structures: helix, single skyrmion, and perfect ferromagnet.
In the right panel of Fig.~3C, we have shown the typical real-space patterns for the helix and the skyrmion, indicating that the skyrmion is of N\'{e}el type as expected in the interface systems~\cite{Fert2013,Heinze2011,Jiang2015,Moreau-Luchaire2016,Wakatsuki2015}.
We also note that the skyrmion is nearly cylindrical, i.e, its radius is almost independent of the layers (see Section B in the Supplementary Materials for details).
The left panel of Fig.~3C shows the stability of the above three magnetic structures, clearly demonstrating 
that the skyrmion state is realized under larger $D$ and smaller $m$ than the ferromagnetic state.
For instance, at $D=0.3$,  stabilized is the cylindrical N\'{e}el-type skyrmion ($m\leq6$) and the ferromanet ($m\geq7$), as depicted in the schematics of Fig.~3A.
This is consistent with the experimental result, the emergence of the THE in smaller $m$ ($m\leq6$) for a given $D$ at the SrRuO$_3$-SrIrO$_3$ interface.
We can thus conclude that the THE in our bilayers stems from the interface DM interaction and the resultant N\'{e}el-type skyrmions.
We also find that the stable regions for the single skyrmion and the ferromagnetic state are divided by an almost linear function of $m$.
This is understood by introducing $D_\mathrm{eff}=D/m$ as follows.
Because of the cylindrical nature of the skyrmion mentioned above, the magnetic moments $\bm{n}_{l,\bm{i}}$ are almost parallel to each other between layers.
The total intralayer ferromagnetic interaction is thus just $m$ times as large as that of the single layer, equivalent to the scale-down of the effective DM interaction given by $D_\mathrm{eff}=D/m$.

With the obtained $D_\mathrm{eff}$, we discuss more detailed energetics in a two-dimensional ferromagnet described by three parameters, $D_\mathrm{eff}$, $J$, and $K$, where $K$ denotes the anisotropy constant. 
Since SrRuO$_3$ is a ferromagnet with perpendicular easy axis, both $J$ and $K$ are positive.
In this situation, it is known that these parameters can be reduced to a single parameter, $\kappa=\sqrt{D_\mathrm{eff}^2/JK}$, to describe the magnetostatics;
the sign of $D_\mathrm{eff}$ does not change the energetics while it determines helicity of the skyrmion~\cite{Bogdanov1989,Bogdanov1994}. 
In the theory considering both the magnetocrystalline anisotropy and the diploar interaction, there is a critical value in $\kappa$ which differentiates the skyrmion phase and the ferromagnetic phase; the former is favored for larger $\kappa$~\cite{Bogdanov1989,Bogdanov1994} corresponding to smaller $m$ in our case.
In SrRuO$_3$, $K$ is reported to be 0.64~Jcm$^{-2}$~\cite{Kanbayasi1976}.
If we assume that bulk SrRuO$_3$ is a three dimensional Heisenberg ferromagnet with $S=1$, $J= (3/2)T_c = 240$~K in the mean-field approximation.
Since the critical value of $\kappa$ is about unity~\cite{Bogdanov1989,Bogdanov1994}, the effective magnitude of DM vector $|D_\mathrm{eff}|$ should be close to $\sqrt{JK} = 2.3$~meV to stabilize the skyrmion state for the $m\leq6$ samples.
The real $|D|$ value defined at the interface is thus deduced to be $m|D_\mathrm{eff}| = 14$~meV.
This is much larger than those in the interface between metals:
$-$2.2~meV for permalloy/Pt~\cite{Nembach2015} and $-$1.05~meV for a Co/Ni multilayer on Pt(111)~\cite{Chen2013}.

In the skyrmion systems, a single skyrmion can be regarded as one flux quantum ($\phi_0 = h/e$, where $h$ is  the Planck constant and $e$ is the elementary charge) in the limit of strong spin-charge coupling.
Then the skyrmion density ($n_\mathrm{sk}$) gives rise to emergent magnetic field ($b$) as $b=n_\mathrm{sk} \phi_0$.
The topological Hall resistivity is hence represented by the following formula:
\[ \rho_H^T = P R_0 b = P R_0 n_\mathrm{sk} \phi_0, \]
where $P$ denotes the spin polarization of conduction electron in SrRuO$_3$.
Since $P$ is found to be $-9.5$\% by the tunneling magnetoresistance in a junction with a 50-nm thick SrRuO$_3$~\cite{Worledge2000}, we can deduce $n_\mathrm{sk}$.
To roughly estimate the separation of the skyrmions, we plotted $n_\mathrm{sk}^{-1/2}$ in Fig.~3B with $\rho_H^T(m)$, the maximum of the topological Hall resistivity in $T$-$H$ plane, and $R_0$ as a function of $m$.
Since there is some uncertainty if the spin polarization value of $-9.5$\% can be applied to our ultrathin SrRuO$_3$ films, we also appended $n_\mathrm{sk}^{-1/2}$ for $P= -1$\% and $-20$\%.
We can approximately estimate that the separation of skyrmions ($n_{sk}^{-1/2}$) is 10--20~nm in our bilayers.
This is an area-averaged value; its decrease with $m$ corresponds to the instability of the skyrmion phase as already discussed.
Considering the above-mentioned coexistence between the skyrmion and the ferromagnetic phase, the local inter-skyrmion spacing can be even smaller.
Nevertheless, the $n_{sk}^{-1/2}$ value of 10--20~nm still provides an indication of the length scale of the skyrmion since its lower limit should be larger than the the film thickness of $\sim$2~nm due to the two-dimensional nature of the skyrmion. 
We also note that this value is comparable to that of the bulk B20 compounds and therefore is promising as well in terms of possible recording memory density~\cite{Koshibae2015}. 
The result indicates that the interface DM interaction at the oxide heterojunction is one of the promising ways to produce skyrmions.
Such functionality is achieved only by the atomically flat oxide interfaces endowed with high tunability.

\section*{Discussion}

We have observed topological Hall effect in the SrRuO$_3$-SrIrO$_3$ bilayers. 
By investigating the ferromagnet thickness dependence, we have demonstrated that the skyrmion phase is driven by the interface DM interaction as a consequence of both the broken inversion symmetry and strong spin-orbit coupling of SrIrO$_3$. 
The result provides a basis for designing and controlling the interface-driven skyrmions in non-chiral magnets, which should be indispensable to future topological electronics.
In order to fully elucidate the nature of the interface DM interaction, further studies on real space
observation of skyrmions~\cite{suppc} and/or direct measurement of the interface DM vector will be useful.
The latter can be performed by Brillouin light scattering or ferromagnetic resonance~~\cite{Di2015,Nembach2015,Cho2015,Stashkevich2015}.

\section*{Materials and Methods}
\paragraph{Sample fabrication}
Epitaxial bilayers consisting of SrRuO$_3$ and SrIrO$_3$ were deposited on SrTiO$_3$(001) substrates by pulsed laser deposition using a KrF excimer laser ($\lambda$= 248 nm) with a fluence of 1--2~J/cm$^2$ on the target surface.
The oxygen partial pressure and deposition temperature were optimized, at 13~Pa and 680$^{\circ}$C for SrRuO$_3$ and 19~Pa and 610$^{\circ}$C for SrIrO$_3$, respectively.
\paragraph{Measurements}
The magnetization data was recorded by SQUID mangetometer with a magnetic field applied normal to the film plane since the magnetic easy axis of SrRuO$_3$ film is perpendicular to the film plane when grown on SrTiO$_3$(001)~\cite{Klein1996}.
Magneto-optic Kerr effect was measured with a laser at 690 nm in polar geometry by using a photoelastic modulator. 
In transport measurements, we applied 3~$\mu$A to a Hall bar with a dimension of 400-$\mu$m width and 1000-$\mu$m length.
This corresponds to the current density of $\sim$3 $\times$ $10^2$ Acm$^{-2}$ in case of $m=4$.
Antisymmetrization has been performed for both the Hall resistivity and the Kerr rotation angle.
Ordinary Hall term has been subtracted from the Hall resistivity by the linear fitting in a higher magnetic field region.

\bibliographystyle{ScienceAdvances}

\begin{thebibliography}{10}
\bibitem{sign}
We can recognize the negative $\rho_H^T$ in the vicinity of coercive field
  $H_c$ at lower temperatures. Please note, however, we have some experimental
  uncertainty there since the dominating anomalous term changes very abruptly
  at $H_c$.
\end{thebibliography}

\newpage
\noindent \textbf{Acknowledgements:} 
The authors thank X.~Z.~Yu and O.~Tretiakov for fruitful discussions.\\
\noindent \textbf{Author Contributions:} JM conceived the research, conducted the sample fabrication and the transport measurements, and wrote the manuscript. NO measured the magneto-optic Kerr effect. KY and FK perfomed the mangetic force microscopy. WK and NN carried out the model calculation. YT and MK supervised the project. All authors discussed the results.\\
\noindent \textbf{Competing Interests:} The authors declare that they have no competing financial interests.\\
\noindent \textbf{Data and materials availability:} Additional data and materials are available online.

\newpage
\renewcommand{\baselinestretch}{1.5}
\begin{figure}[t]
\centering
\includegraphics[scale=.7]{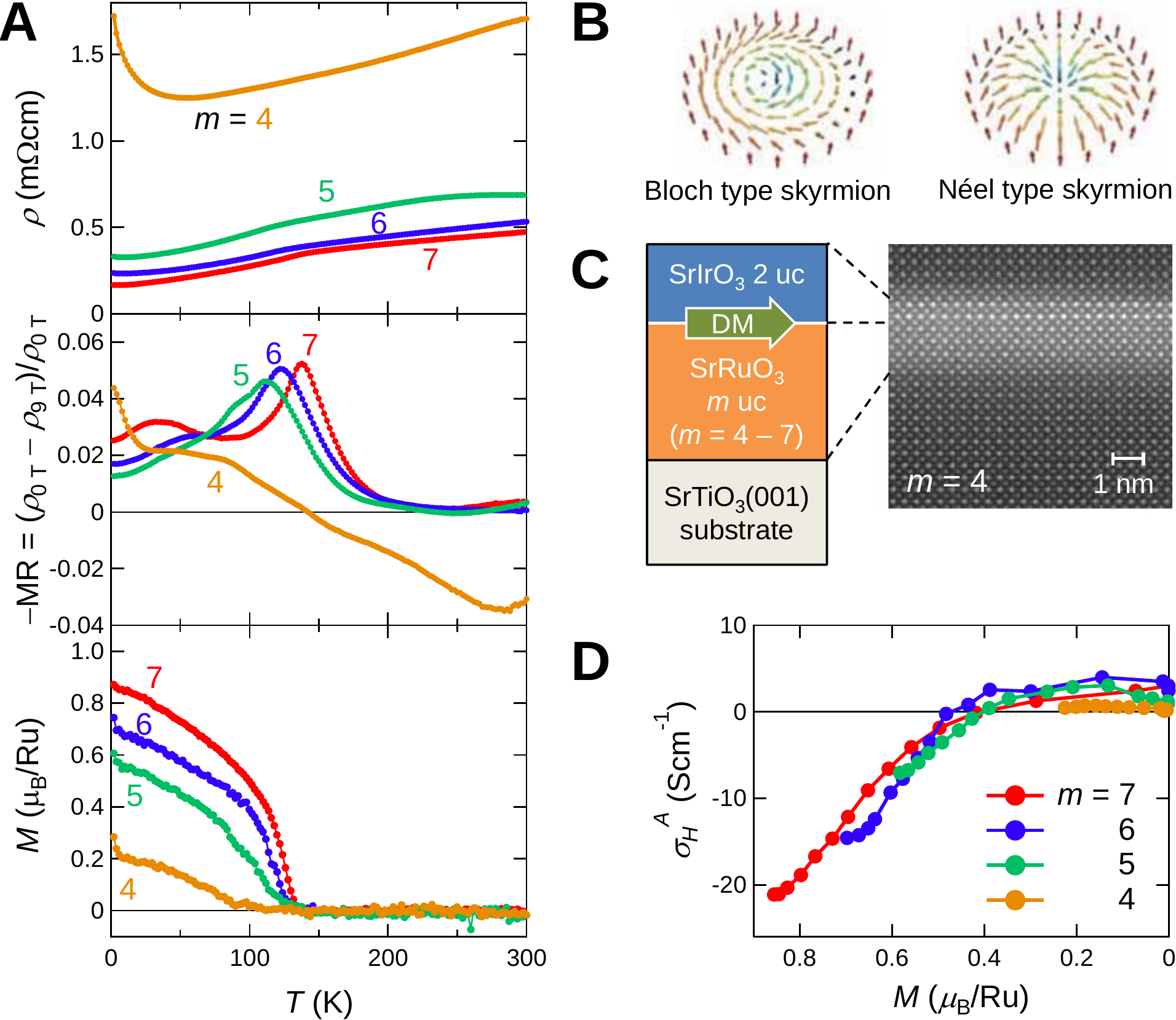}
\caption{\textbf{Structure and basic physical properties of the SrRuO\bm{$_3$}-SrIrO\bm{$_3$} bilayers.}
(\textbf{A}) Temperature ($T$) dependence of resistivity ($\rho$, top panel), magnetoresistance (middle panel), and out-of-plane magnetization measured at 0.05~T (bottom panel) for the (SrRuO$_3$)$_m$-(SrIrO$_3$)$_2$ bilayers ($m = 4$, 5, 6, and 7).
(\textbf{B}) Schematics of Bloch and N\'{e}el type skyrmions.
(\textbf{C}) Schematics and an atomically resolved HAADF-STEM image of the studied bilayer structure.
In the STEM image, SrTiO$_3$ is capped on top of the SrIrO$_3$ layer to protect the surface from electron-beam radiation.
(\textbf{D}) Anomalous Hall conductance ($\sigma_{H}^{A}$) as a function of magnetization ($M$), which was varied via temperarure. 
} 
\label{1}
\end{figure}

\begin{figure}[t]
\centering
\includegraphics[scale=.7]{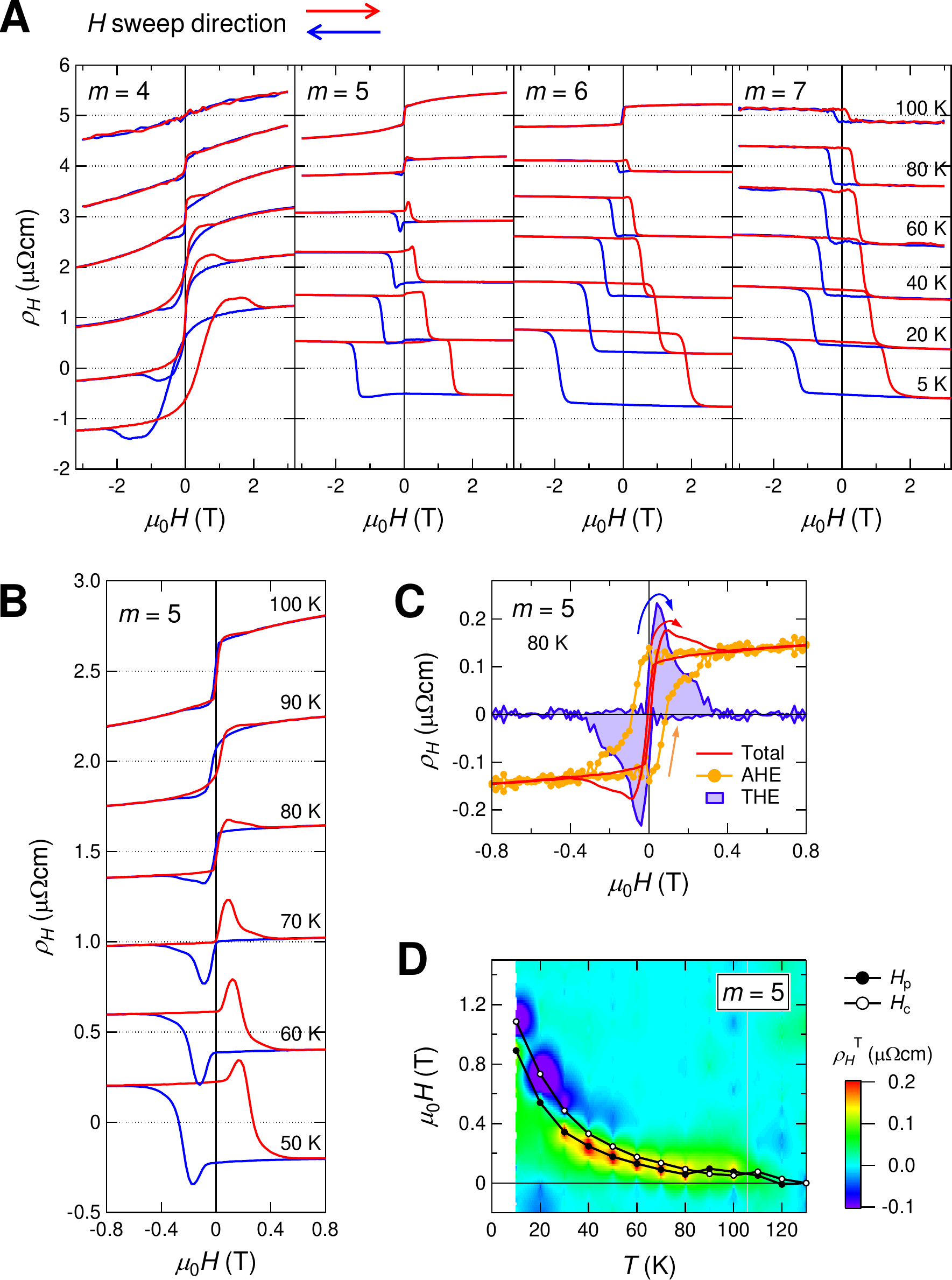}
\caption{\textbf{Hall resistivity of all the bilayers.}
(\textbf{A}) Magnetic-field dependence of Hall resistivity ($\rho_{H}$) of the (SrRuO$_3$)$_m$-(SrIrO$_3$)$_2$ bilayers ($m=4$, 5, 6, and 7) at various temperatures.
Red and blue represent sweep directions of magnetic field.
Ordinary Hall term is subtracted by the linear fitting in a higher magnetic field region.
(\textbf{B}) Detailed view of the Hall resistivity of $m=5$. 
(\textbf{C}) The contribution from anomalous and topological Hall effects of $m=5$ at 80~K (see text for details).
(\textbf{D}) A color map of topological Hall resistivity in the $T$-$H$ plane for $m=5$. Black open and filled symbols represent coercive field ($H_c$) and the field at which topological Hall resistivity reaches its maximum ($H_p$), respectively.
}
\label{2}
\end{figure}

\begin{figure}[t]
\centering
\includegraphics[scale=.7]{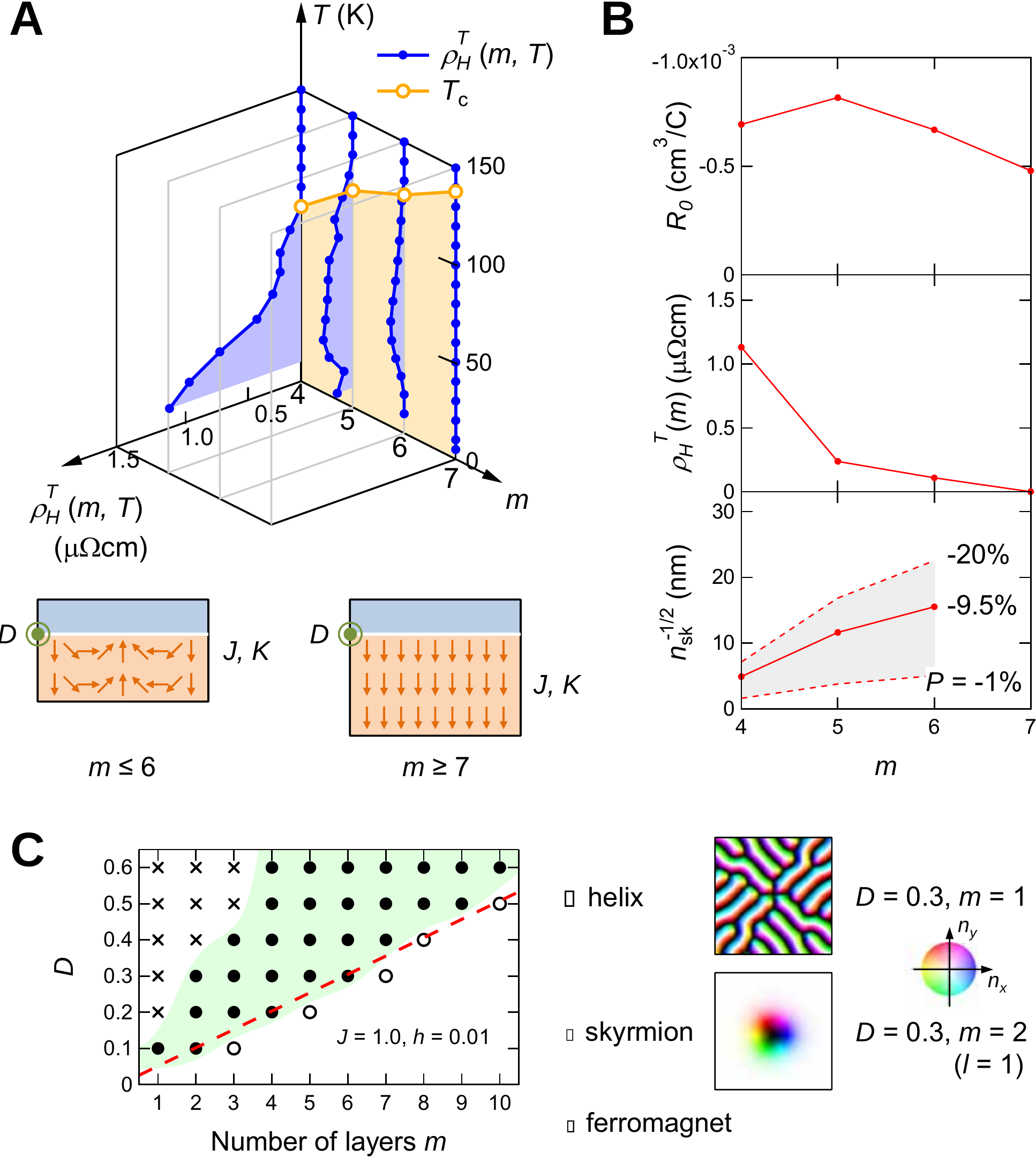} 
\caption{\textbf{Topological Hall effect and calculated stability of skyrmions as a function of the ferromagnet thickness.}
(\textbf{A})
Topological Hall resistivity as functions of $m$ and $T$. Curie temperature ($T_\mathrm{C}$) of the bilayers is also shown.
The schematics below indicate the relationship between the spin structure and interface DM interaction depending on SrRuO$_3$ thickness.
(\textbf{B}) $m$-dependence of the ordinary Hall coefficient ($R_0$, top panel), the maximum of the topological Hall resistivity in $T$-$H$ plane [$\rho_H^T(m)$, middle panel], and the inverse of the square root of the possible skyrmion density ($n_\mathrm{sk}^{-1/2}$ , bottom panel).
(\textbf{C}) Calculated phase diagram of the stable magnetic structures as functions of $m$ and $D$. We have obtained three types of the magnetic structures, namely, helix, skyrmion, and perfect ferromagnet. For the former two, we also show the typical real-space patterns in the right panel. The image size is 150~$\times$~150 unit cells.}
\label{3} 
\end{figure}

\end{document}